\RequirePackage{fix-cm}
\documentclass[twocolumn,epjc3]{svjour3}  
\smartqed  
\RequirePackage{graphicx}
\RequirePackage[numbers,sort&compress]{natbib}
\usepackage{booktabs,verbatim} 
\usepackage[usenames,dvipsnames]{color}
\usepackage{multirow,bigdelim}
\usepackage{ulem}
\usepackage{color}
\usepackage{textcomp} 
\usepackage{tabularx}
\usepackage{amsmath}
\journalname{}
\begin{document}

\title{Measurement of the ambient neutron flux and spectral distribution at the Canfranc Underground Laboratory}

\author{S.~E.~A.~Orrigo\thanksref{e1,IFIC} \and	J.~L.~Tain\thanksref{IFIC} \and	N.~Mont-Geli\thanksref{UPC} \and A.~Tarife{\~n}o-Saldivia\thanksref{IFIC} \and L.~M.~Fraile\thanksref{Madrid} \and M.~Grieger\thanksref{Dresden} \and A.~Quero-Ballesteros\thanksref{Granada} \and J.~Agramunt\thanksref{IFIC} \and	A.~Algora\thanksref{IFIC} \and D.~Bemmerer\thanksref{Dresden} \and F.~Calvi{\~n}o\thanksref{UPC} \and	G.~Cort{\'e}s\thanksref{UPC} \and A.~De~Blas\thanksref{UPC} \and I.~Dillmann\thanksref{Triumf} \and R.~Garc{\'i}a\thanksref{UPC} \and E.~Nacher\thanksref{IFIC}}

\thankstext{e1}{e-mail: Sonja.Orrigo@ific.uv.es (corresponding author)}

\authorrunning{S.~E.~A.~Orrigo $et~al.$} 

\institute{Instituto de F{\'i}sica Corpuscular, CSIC-Universidad de Valencia, E-46071 Valencia, Spain \label{IFIC}
   \and
    Institute of Energy Technologies (INTE), Technical University of Catalonia (UPC), E-08028 Barcelona, Spain \label{UPC}
		\and
	  Grupo de F{\'i}sica Nuclear \& IPARCOS, Universidad Complutense de Madrid, E-28040 Madrid, Spain \label{Madrid}
   \and
	  Helmholtz-Zentrum Dresden-Rossendorf (HZDR), 01328 Dresden, Germany \label{Dresden}   
	 \and
	  Dpto. F{\'i}sica At{\'o}mica, Molecular y Nuclear, Universidad de Granada, Granada, Spain \label{Granada}
	 \and
	  TRIUMF, 4004 Wesbrook Mall, Vancouver, British Columbia V6T 2A3, Canada \label{Triumf}
}
\date{Received: date / Accepted: date} 

\maketitle

\begin{abstract}
We report on the measurement of the ambient neutron flux and its energy distribution over a broad range of neutron energies, conducted in Hall~A of the Canfranc Underground Laboratory using the High Efficiency Neutron Spectrometry Array (HENSA), with a total livetime of 248 days. In particular, we obtained a thermal neutron flux of \mbox{$3.4(2) \times 10^{-6}$ cm$^{-2}$ s$^{-1}$} and a total energy-integrated flux of \mbox{$14.8(2) \times 10^{-6}$ cm$^{-2}$ s$^{-1}$}. This work represents the first long-term measurement of the ambient neutron flux that includes a full spectral characterization across a wide energy interval (from $10^{-4}$ eV to 20 MeV), performed in an underground laboratory to date. The results are relevant for rare-event search experiments located underground, ranging from nuclear astrophysics to astroparticle physics.
\\
\end{abstract}

Experiments aimed at measuring extremely rare events demand an ultra-low background environment and are therefore located in underground laboratories, where the rock overburden screens and reduces background noise from cosmic rays. The flux of muons reaching underground is significantly attenuated compared to the surface; for example, it is reduced by a factor of 10$^5$ \cite{LSC2} at the Canfranc Underground Laboratory (LSC), located at an average depth of 2500 m.w.e. beneath Mount Tobazo in Spain. Consequently, cosmogenic neutrons, produced by cosmic-ray muons, are vastly suppressed underground; however, radiogenic neutrons are generated in the rocks and cavity walls by the spontaneous fission of $^{238}$U and ($\alpha$,n) reactions following the $\alpha$-decay of natural radionuclides. 

Hence neutrons remain a significant source of background for rare-event searches conducted underground. Fast neutrons can undergo elastic scattering, mimicking the signals of Weakly Interacting Massive Particles (WIMPs) sought in dark matter experiments, as well as inelastic scattering (n,n'$\gamma$) producing $\gamma$-rays in the region of interest for neutrinoless $\beta\beta$ decay searches. Low-energy neutrons ($<$1 keV) can undergo capture reactions (n,$\gamma$), further impacting 0$\nu\beta\beta$ measurements and nuclear astrophysics experiments. This multifaceted impact of neutrons on a broad spectrum of underground experiments necessitates a complete characterization of the ambient neutron flux at the experimental location. This is crucial for designing appropriate shielding systems and enhancing the sensitivity of rare-event searches.

We have performed a long-term measurement of the neutron flux in Hall~A of LSC, using the High Efficiency Neutron Spectrometry Array (HENSA) \cite{HENSA1}, with a threefold goal: to study the time evolution of the neutron rate, to measure the integral neutron flux, and to precisely characterize the energy spectrum of the neutron background. In \cite{Orr_2022}, we have presented our results on the long-term evolution of the ambient neutron rate, observed over a period of 412 live days and analyzed in connection with environmental measurements at LSC, also investigating possible seasonal variations that could mimic an annually modulated dark matter signal. We have also shown the importance of continuous monitoring of the neutron background with spectral sensitivity to assess potential impacts on ongoing underground experiments \cite{Orr_2022}. 

In this letter, we present results on the total energy-integrated neutron flux measured in Hall~A of LSC and its energy spectrum from thermal neutrons up to about 20 MeV, collected over 248 live days. Although the neutron flux has been measured in a few underground laboratories (see, e.g., \cite{Best_2016, Ber_2023} and references therein), its energy spectrum remains poorly known or entirely unknown. Nevertheless, this information is crucial: due to the energy dependence of neutron cross-sections, different energy distributions can critically affect each underground experiment differently. Our study represents the first long-term measurement of ambient neutron flux with a spectral distribution covering such a broad range of neutron energies reported so far in an underground laboratory.

HENSA \cite{HENSA1} is an array of $^3$He proportional counters, carefully designed to provide high sensitivity over a wide energy range. Its working principle is that of the Bonner Sphere Spectrometer (BSS) \cite{Thomas_2002} and the configuration used in the Hall~A measurement campaign are described in \cite{Orr_2022}, along with details of the measurements and the subsequent data analysis required to extract the neutron rate and study its time evolution. In particular, the results for the neutron counting rate $n_i$ in each detector $i$ can be found in Table 2 of \cite{Orr_2022}, for each of the measurement periods (\textit{Phases}). In addition, we report here the rates measured in the Cd-lined counter E10, which were not included in the analysis presented in \cite{Orr_2022}: $n_{_{E10}}$ = 0.62(4) and \mbox{$0.65(4) \times 10^{-4}$ s$^{-1}$} in \textit{Phases}~1 and 2, respectively. In the following, we build upon the neutron rate results from \cite{Orr_2022} to determine the ambient neutron flux in Hall~A.

The neutron rate observed in each detector results from the combined effect of the incoming neutron flux $\phi_j$ for energy bin $j$ and the detector response $\epsilon_{ij}$ \cite{Orr_2021}:
\begin{equation}
  n_i = \sum_j \epsilon_{ij} \phi_j \,.
  \label{Eq1}
\end{equation}

\begin{figure}[!t]
  \centering
	\includegraphics[width=1\columnwidth]{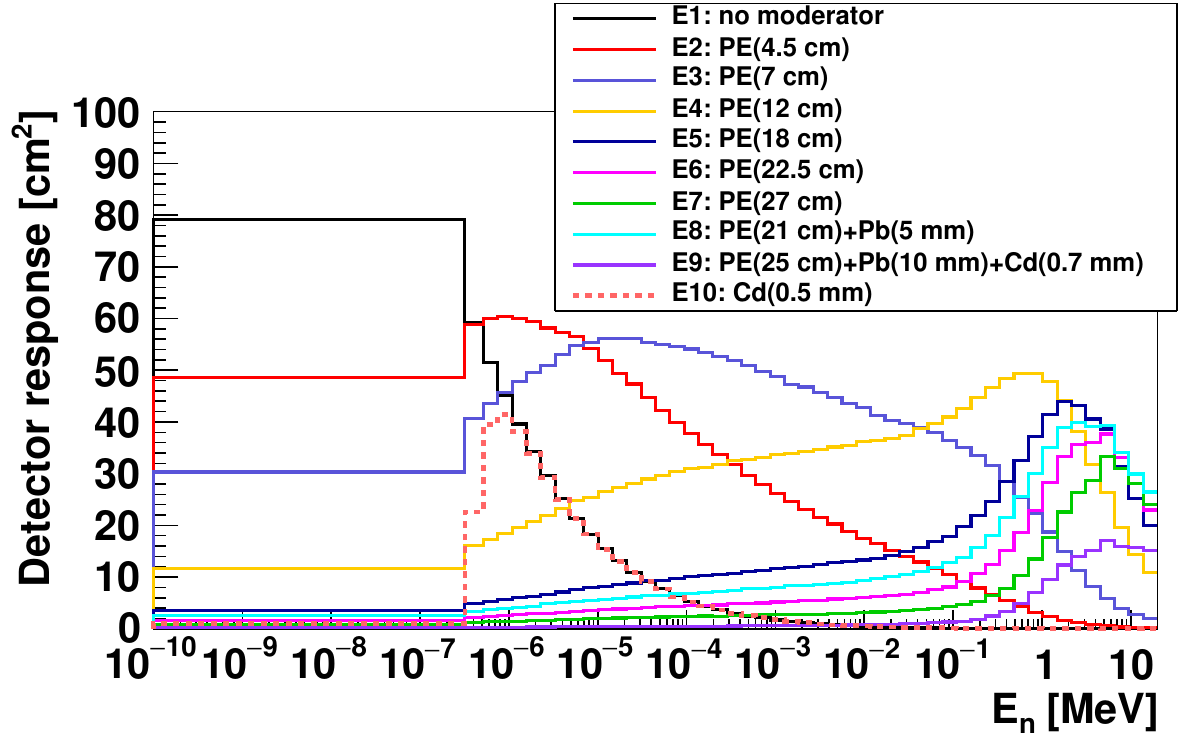}
 	\caption{HENSA response to neutrons simulated with GEANT4. Moderators and covering materials are indicated.}
  \label{Fig1}
\end{figure}

The HENSA response matrix $\epsilon_{ij}$, which reflects the efficiency of each detector as a function of neutron energy $E_n$, was determined by Monte Carlo simulations performed with GEANT4. It is shown in Fig.~\ref{Fig1}, represented in 49 energy bins, with the first thermal bin defined in the interval [10$^{-4}$, 0.32] eV, similar to \cite{Plaza_2023}, followed by 48 equally spaced bins in log($E_n$) up to 20 MeV. The detectors are labeled as in \cite{Orr_2022}, and the thickness of the polyethylene (PE) moderator and the covering materials surrounding the $^3$He counters are also indicated in Fig.~\ref{Fig1}. 

\begin{figure}[!ht]
  \centering
	\includegraphics[width=1\columnwidth]{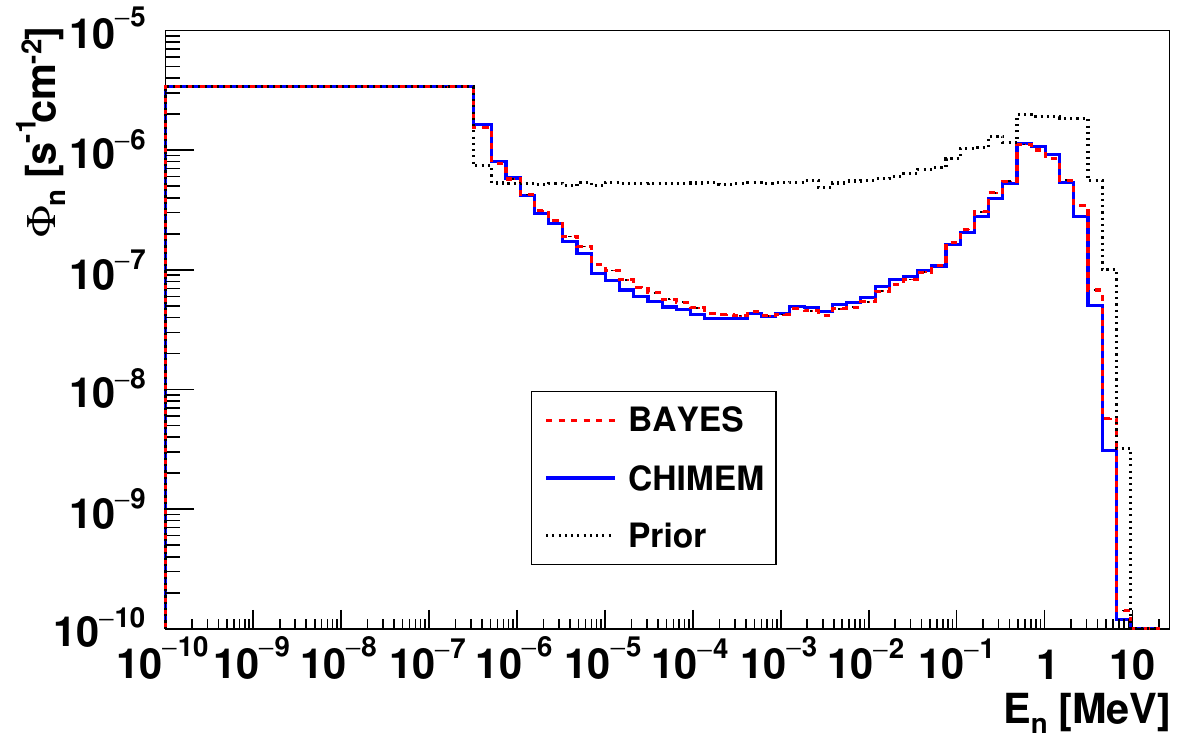}
 	\caption{Energy spectrum of the ambient neutron flux in Hall~A of LSC, obtained by deconvoluting the measured neutron rate \cite{Orr_2022} with the detector response shown in Fig.~\ref{Fig1}, using two different algorithms \cite{Tain_2007}: BAYES (red dashed line) and CHIMEM (blue solid line). The black dotted line represents the prior flux distribution \cite{Mon_2021}, used as the initial solution in both iterative algorithms, with the value in the first energy bin fixed to our determined thermal flux value (see the text).}
  \label{Fig2}
\end{figure}

Given the neutron rate and the response, both the absolute value and spectral shape of the neutron flux can be determined through a deconvolution procedure \cite{Orr_2021,Thomas_2002,Jordan,Grieger_2020} based on iterative algorithms \cite{Tain_2007}. We perform independently the unfolding using two different algorithms: BAYES, which employs the expectation-maximization method, and CHIMEM, based on the maximum-entropy method, both described in \cite{Tain_2007}. The level of agreement between them reflects the quality of the deconvolution and the reliability of the results \cite{Jordan}.

An initial neutron flux distribution (prior information) is typically used in the unfolding procedure to provide the algorithm with a more realistic input than a flat distribution. The prior distribution we use here is represented by the black dotted line in Fig.~\ref{Fig2}, using the same 49 energy bins as for the response (Fig.~\ref{Fig1}). This prior corresponds to the neutron spectrum simulated with FLUKA for Portland concrete in \cite{Mon_2021}, but is recalculated here using an updated version of FLUKA. At LSC, although the rock composition has been investigated \cite{Amare_2006}, the composition of the 40-cm-thick concrete layer, which dominates the final neutron flux, remains unknown. As shown in \cite{Orr_2022}, the neutron rates calculated using the simulated flux curves from \cite{Mon_2021}, which assume different compositions of rock and concrete in Hall~A, reproduce the order of magnitude of our experimental neutron rates. High-energy neutrons are expected to be significantly suppressed at LSC; therefore, the HENSA configuration used during the Hall~A campaign was optimized for sensitivity across a broad neutron energy range, from thermal to $\sim$20 MeV, and lacked efficiency at higher energies \cite{Mon_2023}. Accordingly, the high-energy component was not included in our adopted prior. 

The energy resolving power of our spectrometer, as is the case for any BSS \cite{ReginattoB2002}, is limited, and correlations among the flux values determined in different energy regions may appear. In particular, in the thermal region, it is determined by a single detector (the bare one); thus, the shape of the spectrum below $\sim$0.3 eV cannot be determined. Therefore, we opted to define a single bin covering the entire thermal region, [10$^{-4}$, 0.32] eV, and adopted a different method that allows for an unambiguous determination of the thermal flux, similar to the approach used in \cite{Plaza_2023}. Since the bare neutron counter has maximum sensitivity at thermal energies, while the 0.5-mm-Cd-lined detector is mostly sensitive to the epithermal region (see the responses in Fig.~\ref{Fig1} for E1 and E10, respectively), the thermal neutron flux can be directly determined from the rates measured in these two detectors by solving the system of equations derived from Eq.~\ref{Eq1} for them:
\begin{equation}
 \begin{aligned}
  n_{_{E1}} &= \epsilon_{_{E1,th}} \phi_{th} + \epsilon_{_{E1,ep}} \phi_{ep} \,, \\ 
  n_{_{E10}} &= \epsilon_{_{E10,th}} \phi_{th} + \epsilon_{_{E10,ep}} \phi_{ep} \,.
 \end{aligned}
 \label{Eq2}
\end{equation}

Since $\epsilon_{_{E10,th}}\cong0$ and $\epsilon_{_{E1,ep}}\cong\epsilon_{_{E10,ep}}$, one obtains:
\begin{equation}
  \phi_{th} = {[n_{_{E1}}-n_{_{E10}}]}/{\epsilon_{_{E1,th}}} \,.
  \label{Eq3}
\end{equation}

With this method we obtain a thermal neutron flux of \mbox{$3.4(2) \times 10^{-6}$ cm$^{-2}$ s$^{-1}$}, where the quoted uncertainty is obtained by propagating the statistical uncertainties of the measured rates. Our result is in excellent agreement with the value of \mbox{$3.5(8) \times 10^{-6}$ cm$^{-2}$ s$^{-1}$} measured in \cite{Plaza_2023} at a different location within Hall~A. We then set the value of the first bin in the prior spectrum to the thermal flux value of \mbox{$3.4(2) \times 10^{-6}$ cm$^{-2}$ s$^{-1}$}, determined in this way from Eq.~\ref{Eq3}, and kept this value fixed in the deconvolution procedure.

The energy spectrum of the neutron flux is derived by unfolding the measured counting rates reported in \cite{Orr_2022} with the detector response shown in Fig.~\ref{Fig1}, using the prior information described above and keeping the thermal flux fixed to the value derived from Eq.~\ref{Eq3}. The deconvolution includes the entire period covering \textit{Phases}~1 and 2 (10 months, totaling 248 live days), during which measurements were conducted using the complete HENSA setup with ten detectors \cite{Orr_2022}. It is worth specifying that the Cd-lined counter E10 is used only to determine the thermal flux value, while it is not included in the deconvolution (which uses the other nine detectors), as its inclusion was found to introduce an artificial peak in the spectrum at $\sim$2 $\times 10^{-7}$ MeV. 

For both the BAYES and CHIMEM algorithms, the stopping criterion for the unfolding procedure was set at 300 iterations. This choice was motivated by the observation that, as the number of iterations increased, the reduced chi-square between the experimental data and the reconstructed rates decreased and then stabilized around 300 iterations. Additionally, an excessive number of iterations may lead to overfitting, thereby amplifying fluctuations in the prior spectrum. Hereafter, we present the results obtained using 49 energy bins, which we have verified remain essentially unchanged when using 24 energy bins, as in our initial measurement \cite{Jordan}. 

\begin{figure}[!ht]
  \centering
	\includegraphics[width=1\columnwidth]{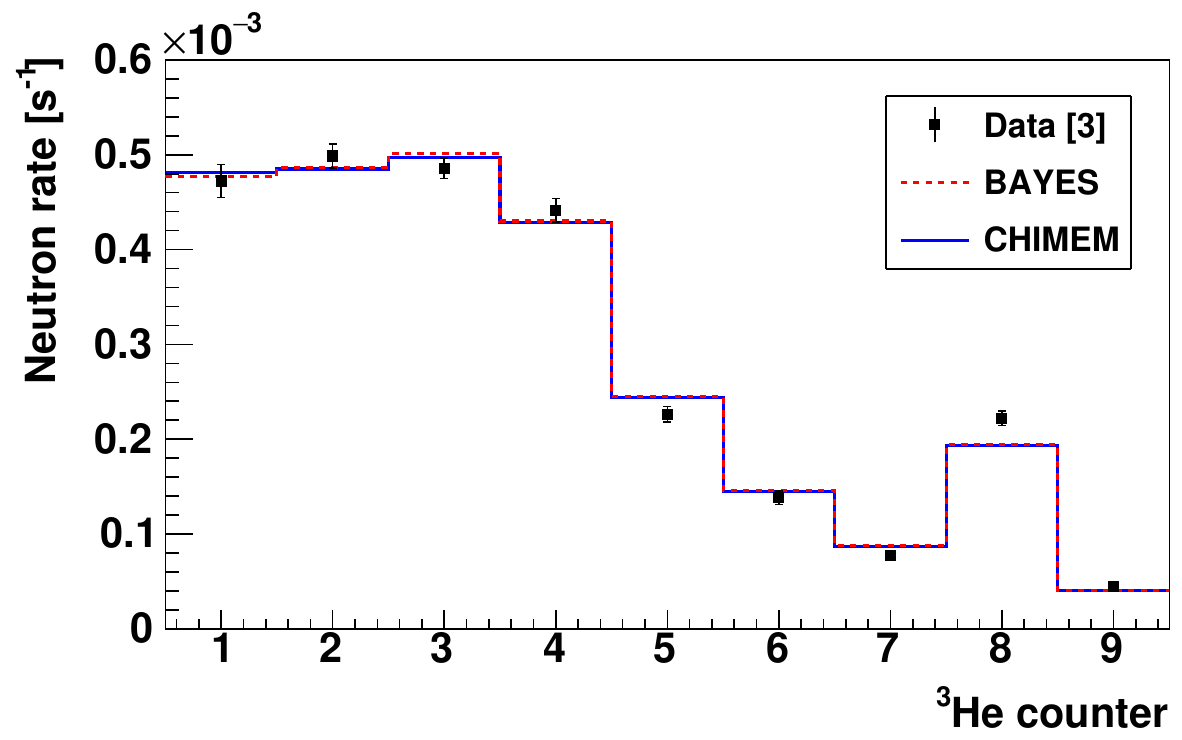}
 	\caption{Reconstructed neutron counting rate using BAYES (red dashed line) and CHIMEM (blue solid line), compared with the data from \cite{Orr_2022} (black points).}
  \label{Fig3}
\end{figure}

The resulting neutron flux distribution as a function of neutron energy is shown in Fig.~\ref{Fig2}, with the red dashed (BAYES) and blue solid (CHIMEM) lines representing the results from the two algorithms. Excellent agreement is observed between them. The reconstructed neutron counting rate obtained using the two different algorithms is compared with the experimental data from \cite{Orr_2022} in Fig.~\ref{Fig3}. Differences between the two algorithms are negligible, whereas they are larger with the experimental rates. This reflects the effect of systematic uncertainties in the deconvolution procedure—coming either from the algorithm, the prior or the calculated response—that prevent it from simultaneously reproducing the rates observed in all detectors. For instance, excluding detector E5 and/or E8 improves the agreement for high-number detectors (corresponding to higher energies) but worsens it for low-number (lower energy) detectors. The level of agreement is similar when using 24 energy bins.

\begin{table}[!t]
	\caption{Neutron flux integrated over different energy regions for the full period covering \textit{Phases}~1 and 2 (10 months, 248 live days). Upper panel: results obtained using the BAYES and CHIMEM algorithms. The thermal flux is fixed to the value determined from Eq. 3, based on the rates measured with the bare and Cd-lined counters (see text). Lower panel: results obtained using the GRAVEL and MAXED algorithms.}
	\label{Tab1}
	\centering
	  \begin{tabular}{llll}
		  \hline
 		  \multicolumn{2}{c}{Neutron energy range (MeV)} & \multicolumn{2}{c}{$\phi_n$ (10$^{-6}$ cm$^{-2}$ s$^{-1}$)} \\ 
 		  \multicolumn{2}{c}{ }                          & BAYES & CHIMEM \\ 
			\hline
			Thermal    & [10$^{-10}$, 3.2$\times 10^{-7}$]   & 3.4(2)$^a$  & 3.4(2)$^a$  \\
			Epithermal & [3.2$\times 10^{-7}$, 0.1]          & 5.96(8) & 5.94(8) \\
			Fast       & [0.1, 20]                           & 5.45(8) & 5.42(9) \\
			Total      & [10$^{-10}$, 20]                    & 14.8(2) & 14.8(2) \\
		  \hline	
 		  \multicolumn{2}{c}{ }                          & GRAVEL & MAXED \\ 
			\hline										
			Thermal    & [10$^{-10}$, 3.2$\times 10^{-7}$]   & 4.2 & 3.0 \\
			Epithermal & [3.2$\times 10^{-7}$, 0.1]          & 5.37 & 5.69 \\
			Fast       & [0.1, 20]                           & 5.34 & 5.98 \\
			Total      & [10$^{-10}$, 20]                    & 14.9 & 14.6 \\
		  \hline	
		\end{tabular}
		\\ \raggedright{$^a$ Fixed to the value determined from Eq.~\ref{Eq3}.}
\end{table}

The upper part of Table~\ref{Tab1} reports the thermal neutron flux determined from Eq.~\ref{Eq3}, as well as the neutron flux obtained by integrating the energy distribution in Fig.~\ref{Fig2} over different energy regions (epithermal, fast and total). The uncertainties reported in the table are statistical and do not include systematic uncertainties arising from the unfolding procedure or the response matrix. They were calculated by propagating the statistical uncertainties in the rates, using the full covariance matrix computed by the BAYES (CHIMEM) code \cite{Tain_2007}. A remarkable agreement is observed between the BAYES and CHIMEM results for the integrated fluxes. We find a total neutron flux of \mbox{$14.8(2) \times 10^{-6}$ cm$^{-2}$ s$^{-1}$} in Hall~A of LSC. This result is in good agreement with the value of \mbox{$13.8(14) \times 10^{-6}$ cm$^{-2}$ s$^{-1}$} obtained in our previous, shorter (25.3 live days) measurement \cite{Jordan}, carried out in 2011 with a reduced setup in an almost empty Hall~A, before the start of any scientific experiment at LSC.  

We emphasize that the results in Table~\ref{Tab1} are robust and practically independent of the chosen prior: performing the deconvolution for the other four concrete compositions of \cite{Mon_2021}, while keeping the thermal flux fixed, we find that the integrated flux in the different regions agrees within the quoted uncertainties for all priors. 

The comparison of results obtained with different codes enables the investigation of systematics associated with the unfolding algorithm \cite{Jordan}. Therefore, in addition to the BAYES and CHIMEM codes, we have analyzed the data using the GRAVEL and MAXED codes, which are commonly employed in the analysis of BSS data. GRAVEL is a non-linear least-squares minimization algorithm based on a modified version of the SAND-II code \cite{Matzke2003}, while MAXED \cite{Reginatto2002} is based on the maximum-entropy principle. The lower part of Table~\ref{Tab1} presents the results obtained with these algorithms, which are consistent with those from BAYES and CHIMEM. The level of agreement with the experimental rates obtained using GRAVEL and MAXED is also similar to that shown in Fig.~\ref{Fig3} for BAYES and CHIMEM, pointing to the hypothesis that this behavior is related to systematic uncertainties in the response determination.

The GRAVEL and MAXED codes currently do not include the option of fixing the thermal flux, which we consider would be very useful if implemented. The thermal flux obtained with GRAVEL is higher than that obtained with MAXED (4.2 and 3.0 \mbox{$\times 10^{-6}$ cm$^{-2}$ s$^{-1}$}, respectively). This variation between the two algorithms is reflected in the flux in the other energy regions, which is slightly smaller in GRAVEL than in MAXED. When the thermal flux is higher, less flux is left for the other regions, and {\it{vice~versa}}. Deriving the thermal flux using an independent method (Eq.~\ref{Eq3}) and then keeping it fixed in the deconvolution provides more stability to the solution in the other energy regions, as seen when comparing BAYES and CHIMEM. The total neutron flux value agrees across all four algorithms.

Another point to consider is the observation reported in \cite{Orr_2022} that, during the first three months of our measurement campaign (October-December 2019), the thermal rate was about 16\% higher than in the subsequent 14 months, likely due to changes in the surrounding environment. The LSC staff informed us that a sizable amount of material—particularly methacrylate and PE, which are effective neutron moderators and can also absorb thermal neutrons to some extent—was added to the middle of Hall~A in December 2019. We also analyzed the time evolution of environmental variables (radon concentration, air temperature, air pressure and humidity), which are regularly measured inside the experimental hall, and found no apparent correlation with this event \cite{Orr_2022}. A similar decrease in the thermal rate between 2019 and 2020 was also independently observed in \cite{Plaza_2023}, whose measurement period partially overlapped with ours. 

Aiming to compare with the results in Table~\ref{Tab1} (10 months), we have repeated the analysis excluding the first three months of \textit{Phase}~1 (i.e., leaving 7 months in total). In this case, we obtain a thermal flux of $2.9(2) \times 10^{-6}$ cm$^{-2}$ s$^{-1}$ from Eq.~\ref{Eq3}. Then, using the BAYES algorithm, we obtain the following values for the neutron flux (in units of 10$^{-6}$ cm$^{-2}$ s$^{-1}$): 6.41(10), 5.09(9) and 14.4(2) for the epithermal, fast and total flux, respectively. Very similar values are obtained using CHIMEM, in the same order and units: 6.59(7), 4.94(7) and 14.4(1). As expected, the variation observed in the neutron rate is reflected in a change in the neutron flux. In particular, compared to Table~\ref{Tab1}, a reduction is observed in the thermal and fast components, as well as in the total flux, while an increase is found in the epithermal contribution. As mentioned above, moderating/absorbing materials were added to Hall~A, so they may reduce the thermal and fast components while shifting a fraction of the fast neutrons into the epithermal range. This finding emphasizes the importance, already highlighted in \cite{Orr_2022}, of continuously monitoring the neutron background and its energy distribution, as the effects of modifications in the hall or variations in meteorological conditions are not easily predictable and may affect the low-rate experiments running underground.

This investigation provides the first comprehensive, long-term measurement of the ambient neutron flux and its energy spectrum over such a broad energy range in an underground laboratory environment. Moreover, together with \cite{Orr_2022}, it highlights the need to maintain permanent setups of $^3$He counters for continuous monitoring of the neutron flux in Halls~A and B at LSC, thereby establishing the basis and demonstrating the feasibility of such long-term measurements.

\bibliographystyle{spphys}
\bibliography{references}

\begin{acknowledgements}
This work was supported by the Spanish Grants No.~CEX2023-001292-S, PID2022-138297NB-C21, PID2019-104714GB-C21, PID2019-104714GB-C22, RTI2018-098868-B-I00 (funded by MCIU, MCIN, MINECO /AEI/FEDER 10.13039/501100011033) and the Generalitat Valenciana Grant No.~PROMETEO/2019/007. We are grateful to Laboratorio Subterr{\'a}neo de Canfranc for hosting the HENSA spectrometer and for the support received by the LSC personnel during the measurement campaign in Hall~A.
\\
\\
\textbf{Data Availability Statement} This manuscript has no associated data or the data will not be deposited. [Authors' comment: The experimental data discussed in this publication are provided in Table~\ref{Tab1}. They were obtained by deconvoluting the neutron-rate data published in \cite{Orr_2022} with the response shown in Fig.~\ref{Fig1}.]
\end{acknowledgements}

\end{document}